\documentclass[12pt]{iopart}

\usepackage{graphicx}
\begin{document}

\title{Quantum mechanics and rational ignorance}

\author{Riccardo Franco
\footnote[3]{To whom correspondence should be addressed
riccardo.franco@csi.it}}
\address{Dipartimento di Fisica and U.d.R. I.N.F.M., Politecnico di Torino
C.so Duca degli Abruzzi 24, I-10129 Torino, Italia
\\
riccardo.franco@polito.it}

\date{\today}

% ----------------------------------------------------------------
\begin{abstract}
In the present article we use the quantum formalism to describe
the process of choice under rational ignorance. We consider as a
basic task a question or an issue where the only answers are 0 and
1. We show that under rational ignorance the opinion state of a
person can be described as a qubit state. We analyze the
predictions that the quantum formalism give in the study of
rational ignorance. We find that answers to different uncorrelated
questions are contextual, that is they are influenced by the
previous questions, even if uncorrelated. Another interesting
prediction is an uncertainty effect which holds when we consider
the statistical variance of two or more questions under rational
ignorance.
\end{abstract}

\pacs{
%03.67.Mn, 03.65.Wj, 42.50.Dv
}
\maketitle

% ----------------------------------------------------------------
%
\section{Introduction}
%%%%%%%%%%%%%%%%%%%%%%%%%%%%%incipit
%%%
Is the microscopic world the only part of  reality which can be
described with the formalism of quantum mechanics? Are there other
phenomena which manifest some \textit{quantum features}? In this
article, we show that the behavior  of people under rational
ignorance can be described within the quantum mechanics formalism.

%%%%%%%%%%%%%%%%%%%%%%%%%%%%%%%%%%%%%%%%%%prima descrizione
%%%
%Rational ignorance is a term most often found in economics,
%particularly in the public choice theory, but also in other
%disciplines which examine rationality and choice.
%
The theory of rational ignorance  was first introduced  by Downs
\cite{Downs1957} to study why voters know so little about
important issues of politic life, and it is based on the following
statement: when the expected benefits of information are too small
relative to the costs (for example in an election), people in
general choose to remain uninformed.
In other words, we have rational ignorance about an issue when any
potential benefit deriving from an informed decision about that
issue can outweigh the cost of acquiring enough information to
make the decision. This has consequences for the decisions made by
large numbers of people, such as general elections, where for each
person the probability of changing the outcome with one vote is
very small.

%
%%%%%%%%%%%%%%%%%%%%%%%%%%%%%%%%%%%%%%%%%%%%%%%%%%%%%%%%%%%%%%%%%%%%%%%%
%%%%scopo e risultati
In this article, we will consider only questions or issues for
which the rational ignorance is the dominant process of choice: we
also call this situation a rational-ignorance regime. In such a
situation, we will show that the quantum formalism can be used to
describe simple questions or issues where the only answers are 0
or 1. We will introduce the general formalism of quantum mechanics
and in particular some simple notions of quantum information
theory, such as the qubit.

The main results of this article are: 1) the opinion-state of a
person can be represented, in rational-ignorance regime, by a
qubit state; 2) the different issues or questions in
rational-ignorance regime can be written as operators acting on
the Hilbert space of the states 3) the explicit answer of a person
to a question in regime of rational ignorance can be described as
a collapse of the opinion vector onto an eigenvector of the
corresponding operator 4) there is a contextuality effect, which
influences the opinion-state in the case of repeated  issues or
questions 5) in rational-ignorance regime an uncertainty principle
holds, which states that the sum of the variances relevant to at
least two questions has a non-trivial lower bound.

It must be noted that our formalism will only describe the
behavior under rational ignorance, without considering the
psycological motivations that lead to this phenomenon.
%
%
%%%%%%%%%%%%%%%%%%%%%%%%%%%%%%%%%%%%%%%%%%%%%%%%%%%%%%%%%%%%%%%%%%%%%%%%
%%%%organizzazione
This article is organized as follows: in section \ref{formalism
and test} we introduce the basic notation of quantum mechanics and
we describe the answers to a question or an issue in
rational-ignorance regime in terms of bra and ket. In section
\ref{collapse} we define some important operators which are useful
to describe a question, and we introduce the concept of state
vector after an answer. In section \ref{lur} we write the
generalized uncertainty principle for questions under
rational-ignorance regime.
\section{Quantum formalism and basic tests}\label{formalism and test}
%%%%%%%%%%%%%%%%%%%%%%%%%%%%%%%%%%%%%%%%%%%%%%%%%%%%%%%%%%%%%%%%%%%%%
%%Formalismo
%%%%%%%%%%%%%%%%%%%%%%%%%%%%%%%%%%%%%%%%%%%%%%%%%%%%%%%%%%%%%%%%%%%%%
We briefly introduce the standard bra-ket notation usually used in
quantum mechanics, introduced by Dirac \cite{Dirac}. The state of
a quantum system is identified with a unit ray in a complex
separable Hilbert space, called ket $|s\rangle$. When the
considered Hilbert space is finite-dimensional, the ket can be
written, given an orthonormal basis, as a column vector
$(c_1,c_2,c_3,...)^T$, while the dual vector to the ket, called
the bra $\langle s|$, can be written, given the same orthonormal
basis, as a row vector $(c_1^*,c_2^*,c_3^*,...)$. The inner
product, defined in the Hilbert space, between bra $|s\rangle$ and
ket $\langle s'|$ can be written as a bracket $\langle
s'|s\rangle$, thus giving a complex number, called probability
amplitude. In finite dimensional Hilbert space, $|\langle
s'|s\rangle|^2$ is the probability to get the state $|s'\rangle$
as the result of an opportune measure on the state $|s\rangle$.

The simplest quantum system is given by a vector in a discrete Hilbert space with dimension 2: this system represents
the unit of quantum information, or qubit. Given a basis in the Hilbert space, we can define two vectors
$$
|0\rangle = \left (
  \begin{tabular}{c}
    1 \\
    0
    \end{tabular} \right ) \,,\,\,
|1\rangle = \left (
  \begin{tabular}{c}
    0 \\
    1
    \end{tabular} \right )\,\,,
$$
representing the quantum analogue to the two possible values 0 and
1 of a classical bit. An important difference is that in the
quantum case a state can be in a linear superposition of 0 and 1,
that is $a|0\rangle + b |1\rangle$, with $a$ and $b$ complex
numbers.

In quantum mechanics, any observable quantity is described by an
hermitian operator (for example, $\widehat{O}$). Its eigenvalues
$\{o_i\}$ represent the measurable values of that observable,
while its eigenvectors $\{|o_i\rangle\}$ the corresponding quantum
states. In the case of a qubit, there are three important
observables, connected with the Pauli matrices:
\[
 \begin{tabular}{ccc}
  $\widehat{\sigma}_x=\left[
    \begin{tabular}{cc}
    0 & 1 \\
    1 & 0
    \end{tabular}\right]$
       &
  $\widehat{\sigma}_y=\left[
    \begin{tabular}{cc}
    0 & $-i$ \\
    $i$ & 0
    \end{tabular}\right]$
       &
  $\widehat{\sigma}_z=\left[
    \begin{tabular}{cc}
    1 & 0 \\
    0 & -1
    \end{tabular}\right]$ \, .
 \end{tabular}
\]
The eigenvalues of these hermitian operators define three
orthonormal basis, with the peculiarity that each eigenvector of
one basis is an equal superposition of the eigenvectors of any of
the other basis; such basis are also called mutually unbiased:
\begin{equation}\label{unbiased}
\{|0\rangle, |1\rangle\}, \left\{\frac{|0\rangle \pm
|1\rangle}{\sqrt{2}}\right\},  \left\{\frac{|0\rangle \pm i
|1\rangle}{\sqrt{2}}\right\}
\end{equation}
In the microscopic world, a physical realization of a qubit is
provided by a spin $1/2$ particle, like an electron. We can
measure the spin along the directions $x$, $y$ and $z$, with
results $\pm 1/2$. These measurements are connected to the
observables $\widehat{\sigma}_x$, $\widehat{\sigma}_y$,
$\widehat{\sigma}_z$.

It is important to recall that two operators $\widehat{A}$ and
$\widehat{B}$ are called commuting where
$[\widehat{A},\widehat{B}]=\widehat{A}\widehat{B}-\widehat{B}\widehat{A}=0$,
and that the Pauli matrices are non-commuting operators. In
quantum mechanics it is impossible to measure contemporarily two
observables with non-commuting operators.

%%%%%%%%%%%%%%%%%%%%%%%%%%%%%%%%%%%%%%%%%%%%%%%%%%%%%%%%%%%%%%%%%%%%%
%%rational ignorance
%%%%%%%%%%%%%%%%%%%%%%%%%%%%%%%%%%%%%%%%%%%%%%%%%%%%%%%%%%%%%%%%%%%%%
In the context of rational ignorance, what are the observable
quantities?  As a basic test, we can observe the answer of people
to a precise question or issue, in a situation where the costs (in
terms of time and other resources) to get the correct answer
outweighs the expected benefits. In particular, we will consider
only questions for which the possible answers can only be 0 or 1
(false or true). We can perform on the same person different basic
tests, that is we can ask different questions  or issues in regime
of rational ignorance, in order to know to opinion state of the
person.

Given a question or an issue, let us associate to the answers 0
and 1 (false and true respectively) the vectors $|0\rangle$ and
$|1\rangle$, corresponding to the truth values of the answer.
Since the Pauli matrices have all eigenvalues $\pm 1/2$, an
important operator is the projector
$\widehat{P}_z=(I-\widehat{\sigma}_z)/2$, whose eigenvalues are
$0,1$ and whose eigenvectors are $|0\rangle$ and $|1\rangle$. Thus
$\widehat{P}_z$ can be considered the hermitian operator (or
observable) associated to a question in context of rational
ignorance.

Let us now consider a second question, with the important
assumption that the answers to question 1 and 2 are statistically
independent: for any answer to question 1, there is the same
probability that the answer to question 2 is 0 or 1.  This means
that we must consider the mutually unbiased basis of formula
(\ref{unbiased}), and the second question can be associated only
to observable $\widehat{P}_x=(I-\widehat{\sigma}_x)/2$ or
$\widehat{P}_y=(I-\widehat{\sigma}_y)/2$.

For example in the case of $\widehat{P}_x$, the answer to question
2 (0 or 1) can be made in correspondence to vectors $(|0\rangle
\pm |1\rangle)/\sqrt{2}$.
\section{Quantum collapse and answers}\label{collapse}
One of the axioms \cite{Dirac} of quantum mechanics states that,
given an initial state $|s\rangle$ in a discrete Hilbert state and
an observable $O$ (with the discrete sets of eigenvalues $\{o_i\}$
and eigenvectors $\{|o_i\rangle\}$), if we measure one eigenvalue
$o_i$ than the resulting state is $|o_i\rangle$. This sudden
change of the state vector is called quantum collapse, and its
consequences are very important in quantum mechanics.

In the context of rational ignorance, the collapse of the state
vector simply states that, given an initial opinion-state
$|s\rangle$ and a question $\widehat{P}_z$, if the answer is 0 the
resulting state is $|0\rangle$, while if the answer is 1 the final
state is $|1\rangle$.

Very simple, isn't it? But let us now give to the same person a second question $\widehat{P}_x$, again in the context
of rational ignorance. Recalling that the inner product between $|0\rangle$ and $(|0\rangle \pm |1\rangle)/\sqrt{2}$ is
$1/\sqrt{2}$, we deduce that answers 0 and 1 for the second question have the same probability 1/2. Thus, when the
person answers to the last question, the state vector $|0\rangle$ is subjected to a second collapse, giving for example
$(|0\rangle - |1\rangle)/\sqrt{2}$ for the answer 0. But if now we ask again the question 1 to the same person, we have
$1/2$ probability to get as answer 1, which is different from before! The rational ignorance situation has manifested
an irrational behavior of the person. The question 1 has been asked two times, but what has been changed is the
context.

The paradoxical situation is due to the usual belief that we can
assign pre-defined elements of reality to individual observables
also in regime of rational ignorance. For example, if we ask to a
person the three questions associated to observables
$\widehat{P}_x$, $\widehat{P}_y$, $\widehat{P}_z$, we can write
three-elements arrays $(*,*,*)$ , with $*$ is $0$ or $1$,
corresponding to 8 the combinations of possible answers. The
opinion state of each person can be classically associated to one
of these combinations, for example the array $(1,0,1)$, and any
repetition of question $i-th$ will give the $i$-th element of the
array, independently form the previous answers. This situation,
analogue to classic mechanics, is true when the expected benefits
of information are high relative to the costs of getting
information. In a rational ignorance regime, instead, we cannot
write arrays of answers like $(1,0,1)$ if the related observables
are non-commuting. We say that the answers to these questions can
not be known contemporarily, thus giving an important limit to the
complete knowledge of the opinion state of a person in
rational-ignorance regime.

This effect in microscopic world is called quantum contextuality \cite{contextuality}, and evidences, for any
measurement,  the influence of other non-commuting observables previously measured. The repeated sequence of questions
in regime of rational ignorance is the analogue of the repeated Stern-Gerlach experiments in quantum mechanics.

\section{The uncertainty relations and rational ignorance}\label{lur}
In quantum physics, the Heisenberg uncertainty principle is a
mathematical limit on the accuracy with which it is possible to
measure everything there is to know about a physical system. In
its simplest form, it applies to the position and momentum of a
single particle in a mono-dimensional continuous space, but more
general definitions have given in \cite{Simon_99}.  In Hilbert
spaces with discrete dimension have been formulated other new
general forms of uncertainty relations \cite{lur}.

The uncertainty of an observable $\widehat{A}$ for any given
quantum state is defined as the statistical variance of the
randomly fluctuating measurement outcomes
$$
\delta^2(\widehat{A})=\langle \widehat{A}^2 \rangle - \langle
\widehat{A} \rangle ^2
$$
Variance zero means that in any experimental measure of the
observable we always obtain the same result.

The local uncertainty relations \cite{lur} state that, for any set
$\{\widehat{A},\widehat{B},\widehat{C},...\}$ of non-commuting
operators, there exist a non-trivial limit $U$ such that
$\delta^2(\widehat{A})+\delta^2(\widehat{B})+\delta^2(\widehat{C})+...
\geq U$. In the case of projectors $\widehat{P}_x, \widehat{P}_y,
\widehat{P}_z$ we have
\begin{equation}\label{lur_eq}
\delta^2(\widehat{P}_x)+\delta^2(\widehat{P}_y)+\delta^2(\widehat{P}_z)
\geq \frac{1}{2}
\end{equation}

In terms of rational invariance, formula (\ref{lur_eq}) states
that given three uncorrelated questions in regime of rational
ignorance, the sum of their statistical variances have a lower
non-null bound. We remember that zero variance for one observable
means that for every person the answer to the question is the same
(complete agreement). Thus equation (\ref{lur_eq}) states that
there is a natural limit to the sum of the variances for questions
under rational ignorance.
\section{Conclusions}\label{conclusions}
In the present article, we have given a mathematical description
in terms of quantum formalism of questions or issues in situations
where the rational ignorance is the dominant process of choice
between different answers. The opinion state is thus described in
terms of a state vector in a complex Hilbert space. We have used
three projection operators, deriving from the Pauli matrices and
corresponding to uncorrelated questions: these operators have
unbiased basis vectors. We have shown that the hypothesis of
rational ignorance leads to a direct influence on the answers to
different sequential questions, like the repeated Stern-Gerlach
experiments. This example shows that also the opinion state in
rational ignorance regime is contextual. A correlated result is
finally the uncertainty principle, which evidences a limit also in
the statistical  knowledge  of the opinion state.

This article has the main scope to define a simple formalism to describe rational ignorance, without exploring all the
consequences of such an approach. For example, it remains to explore the effects of rational ignorance on the
communication process between two persons: we expect to evidence some EPR effects, and to find the concept of
entanglement. This study will be presented in a separate paper. Finally, we think that the concept of rational
ignorance can be also applied in the study of practical reasoning, such as the association of contexts to words. There
are many new papers about this topic, which consider the possible contexts of a word in terms of arrays \cite{aerts1,
aerts2, bruza}: in particular a similarity with the quantum mechanics formalism and with the quantum logic is
suggested. What is not clearly considered in these papers is a formal description of contextuality effects which the
quantum measurement theory evidences, and whose consequences remain to be shown. It is quite paradoxical that in the
study of the contexts of a word the problem of contextuality has not yet been fully considered.

 \footnotesize
%
%%%%%%%%%%%%%%%%%%%%%%%%%%%%%%%%%%%%%%%%%%%%%%%%%%%%%%%%%%%%%%%%%%%%%%%%%%%%%%%%
%

\end{document}